\documentclass[aps,prl,twocolumn,superscriptaddress,longbibliography]{revtex4-1}
\usepackage{amssymb} 
\usepackage{graphicx,color} 
\usepackage{bbm} 
\usepackage{multirow}
\usepackage{amsmath}
\usepackage{mathrsfs} 
\usepackage{hyperref}
\usepackage[usenames,dvipsnames]{xcolor}
\usepackage{graphicx}
\usepackage{cancel}
\usepackage{color} 
\definecolor{darkgreen}{rgb}{0,0.6,0}
\definecolor{darkblue}{rgb}{0,0,0.6}
\definecolor{darkred}{rgb}{0.6,0,0}
\definecolor{darkpurple}{rgb}{0.5,0,0.5}

\newcommand{\Weng}{\mathcal{W}^\mathrm{engine}}

\hypersetup{
bookmarksopen=true,
pdftitle="",
pdfauthor="", 
pdftoolbar=false, 
pdfstartview={FitH},		
pdfmenubar=true,			
pdfhighlight=/O,			
colorlinks=true,			
urlcolor=darkblue,
citecolor=darkblue,		
linkcolor=darkblue}

\newcommand{\plaind}{\mathrm{d}}

\newcommand{\ave}[1]{\left\langle#1\right\rangle}
\renewcommand{\exp}[1]{\mathchoice{\mathrm{e}^{#1}}{\operatorname{exp}\left(#1\right)}{\operatorname{exp}\left(#1\right)}{\operatorname{exp}\left(#1\right)}}
\newcommand{\tf}{t_\text{f}}
\newcommand{\lambdaf}{\lambda_\text{f}}

\newcommand{\lambdaopt}{\lambda(t)}
\newcommand{\lambdaoptv}{\lambda_{v_0}(t)}

\newcommand{\xaveopt}{\ave{x(t)}}

\newcommand{\Waveopt}{\ave{W}}

\newcommand{\drift}{\omega}

\DeclareMathOperator{\Var}{Var}

\newcommand{\vgt}{v_0^\mathrm{r}}
\newcommand{\vm}{v_0^\mathrm{m}}

\newcommand{\lambdaengine}{\lambda^\mathrm{engine}}

\newcommand{\elabel}[1]{\label{eq:#1}}
\newcommand{\eref}[1]{(\ref{eq:#1})}
\newcommand{\Eref}[1]{Eq.~(\ref{eq:#1})}
\newcommand{\Erefs}[1]{Eqs.~(\ref{eq:#1})}

\newcommand{\op}{\mathrm{ss}}

\graphicspath{{figures/}}

\allowdisplaybreaks

\newcommand{\Title}{Optimal closed-loop control of active particles and a minimal information engine}

\begin{document}
\title{\Title
}
\author{Rosalba Garcia-Millan}%
\thanks{These authors contributed equally.}
\affiliation{%
Department of Mathematics, King's College London, London WC2R 2LS, United Kingdom}
\affiliation{DAMTP, Centre for Mathematical Sciences, University of Cambridge, Cambridge CB3 0WA, UK}
\affiliation{St John's College, Cambridge CB2 1TP, UK}%

\author{Janik Sch\"uttler}%
\thanks{These authors contributed equally.}
\affiliation{%
DAMTP, Centre for Mathematical Sciences, University of Cambridge, Cambridge CB3 0WA, UK}%
\thanks{These authors contributed equally.}

\author{Michael E.~Cates}%
\affiliation{%
DAMTP, Centre for Mathematical Sciences, University of Cambridge, Cambridge CB3 0WA, UK}%

\author{Sarah A.~M.~Loos}%
 \email{sl2127@cam.ac.uk}
\affiliation{%
DAMTP, Centre for Mathematical Sciences, University of Cambridge, Cambridge CB3 0WA, UK}

\begin{abstract}
We study the elementary problem of moving an active particle by a trap with minimum work input. 
We show analytically that (open-loop) optimal protocols are not affected by activity, but work fluctuations are always increased. For closed-loop protocols, which rely on initial measurements of the self-propulsion, the average work has a minimum for a finite persistence time. Using these insights, we derive an optimal periodic active information engine, which is found to have higher precision and information efficiency when operated with a run-and-tumble particle than for an active Ornstein-Uhlenbeck particle and, we argue, than for any other type of active particle.
\end{abstract}
\maketitle

Exploring strategies to control the individual or collective motion of self-propelled particles~\cite{schneider2019optimal,falk2021learning,liebchen2019optimal,baldovin2023,shankar2022optimal} enhances our understanding of active dynamics 
and represents a crucial step towards  
technological applications, ranging
from active metamaterials to nanosized robots. 
Recent research has {investigated} various means of control, 
such as adjusting the potential landscape~\cite{davis2024active,gupta2023efficient,aurell2011optimal,ekeh2020thermodynamic,majumdar2022exactly,blaber2023optimal,lee2023nonequilibrium}, confinement \cite{malgaretti2022szilard,cocconi2023optimal,casert2024learning}, 
or the level of intrinsic activity \cite{falk2021learning}. 
In the realm of small-scale organisms and machines, high frictional losses and the prevalence of (non-)thermal fluctuations become particularly important. In this context, ``thermodynamically optimal'' control schemes minimize the heat dissipation, work input, or entropy production {associated with} transitioning a system from one state to another. 
Even for passive systems, {such schemes} can {lead} to non-intuitive solutions {with} driving {discontinuities}~\cite{schmiedl2007optimal,blaber2021steps,engel2023optimal,whitelam2023demon,loos2023universal}.
To explore the principles of thermodynamically optimal control of active systems, which are characterized by noise, memory, (non-)Gaussian statistics and intrinsic driving, it is thus  vital to study {paradigmatic} simplified models.

In this spirit, we investigate the elementary problem of 
{dragging} a harmonic trap containing a single active particle 
to a target position within a given finite time, through some fluid {forming} a heat bath, in one spatial dimension. We optimize the dragging protocol to minimize the experimenter's work input, a directly measurable quantity \cite{loos2023universal,ciliberto2017experiments} that bounds below the controller's total energy expenditure.
The problem is complex enough to show key features of optimal control of active matter, yet can be solved exactly. Different from passive particles (PPs), for which this problem was solved in the seminal paper~\cite{schmiedl2007optimal}, self-propelled particles can actively swim towards the target, suggest{ing} that less external work input is needed.  
However, {the} internal self-propulsion can just as well increase the resistance against the translation. We explore under which conditions the intrinsic activity facilitates or hinders the transport. To gain {broader} insights, we consider two models: run-and-tumble particles (RTPs) and active Ornstein-Uhlenbeck particles (AOUPs). 

{From a control theory perspective, the dragging problem represents an} \textit{open-loop} control {scheme}, because the protocol is predefined and fixed. More {sophisticated} are \textit{closed-loop} or feedback control schemes, which dynamically adapt the protocol taking into account real-time knowledge of the {stochastic} system state, resembling a mesoscopic ``Maxwell demon'' \cite{ciliberto2017experiments}. Closed-loop control can generally {provide} more precise manipulation, but it necessarily relies on measurements which, according to Landauer's principle, {have a} thermodynamic cost~\cite{parrondo2015thermodynamics,jun2014high,berut2012experimental,ribezzi2019large}. {As a simple example of closed-loop control,}  we study how an initial measurement of the system state affects the optimal protocol, generalizing previous studies on PPs \cite{abreu2011extracting,whitelam2023demon} to the active case. 
Remarkably, we find that if the particle is initially moving towards the target position, the optimal protocol can involve an initial jump of the trap \emph{away} from the target, challenging naive intuition. 

As an application, we theoretically construct a minimal, thermodynamically optimized iteration that uses measurement information and returns extractable work, creating an ``information engine''~\cite{saha2021maximizing,parrondo2015thermodynamics,saha2023information}. The harvested work stems from the in-built activity~\cite{malgaretti2022szilard,cocconi2023optimal,cocconi2024efficiency,saha2023information}, as in an active heat engine~\cite{krishnamurthy2016micrometre,holubec2020active,fodor2021active,speck2022efficiency,zakine2017stochastic}.

{Our approach complements recent perturbative methods based on response theory \cite{davis2024active,gupta2023efficient}, which enable the study of more complex active systems, but are limited to the weak and slow driving regime. Our exact results allow for an explicit analysis of fast and intermediate driving, where driving discontinuities have been observed in various contexts for PPs~\cite{schmiedl2007optimal,blaber2021steps,engel2023optimal,whitelam2023demon,loos2023universal}.}

\begin{figure*}
\includegraphics[width=\textwidth]{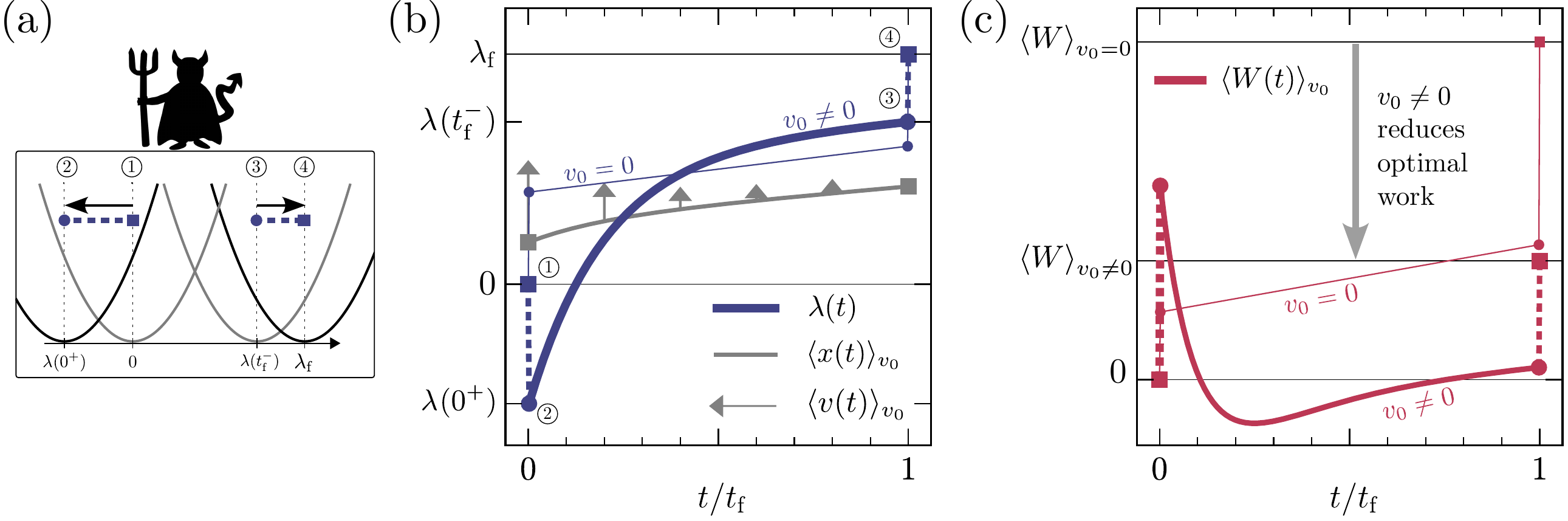}
\caption{
Optimal dragging {for open-loop control and} for closed-loop control, where the demon measures the particle's initial self-propulsion {(here, $v_0>0$ as an example)}. (a) Protocol stages ($1\to2$ initial jump, $2\to3$ continuous driving regime, $3\to4$ final jump to target). (b) Protocol (blue), average particle position (grey) and self-propulsion (arrows). {The thin blue line shows the open-loop protocol, the thick line shows the corresponding closed-loop protocol}. 
(c) \textit{Cumulative} average work $\ave{W(t)}_{v_0} = \int_0^t\plaind{t'}\  \dot{\lambda}k(\lambda - \langle x\rangle_{v_0})$ {for open-loop control (thin line) and for closed-loop control (thick line).}
Here, $\tau = 0.2$, $k = 1$, $\tf = 1$, and $v_0 = 1$.
}
\label{fig:work_incs}
\end{figure*}

\textit{Model.---}We consider a one-dimensional model for a self-propelled particle at position $x$ in a harmonic potential $V(x,\lambda)=k (x-\lambda)^2 /2$ {centered at $\lambda$ with stiffness $k$.}
The particle's motion follows the
overdamped Langevin equation
\begin{equation}
\elabel{oLE}
      \dot{x}(t) = -\partial_x V(x,\lambda) + v(t) + \sqrt{2D} \ \xi(t) \ ,
\end{equation}
where $v$ denotes the self-propulsion,
$D$ is {the} thermal diffusion constant,
and $\xi$ is a thermal, unit-variance Gaussian white noise, $\ave{\xi(t)} = 0$, 
$\ave{\xi(t)\xi(t')} = \delta(t-t')$. 
We have absorbed the friction constant and $k_\mathrm{B}$ into other parameters.
We consider two standard models of active motility: (i)
AOUPs,
where
$ \tau \dot{v}(t) = - v(t) +  \sqrt{2D_v} \ \xi_v(t),$
with self-propulsion ``diffusion'' constant $D_v$ and unit-variance Gaussian white noise $\xi_v$ independent of $\xi$;
and (ii) RTPs, 
where {$v$}
is a dichotomous Poissonian noise
that randomly reassigns values $\pm {\drift}$ with tumbling rate $1/\tau$ \cite{schnitzer1990strategies,schnitzer1993theory,schneider2019optimal,cates2012diffusive,garcia2021run}.
A direct comparison can be established up to second-order correlators
by
fixing ${D_v}/{\tau}\equiv {\drift}^2 $, such that both {models yield}
\begin{equation}
      \ave{v(t)v(t')}_{\op} = {\drift}^2\exp{-\frac{|t-t'|}{\tau}}
\end{equation}
at stationarity (ss), with correlation time $\tau$. 
Comparing AOUPs versus RTPs allows us to study the effect of non-Gaussianity (RTPs) \cite{lee2022effects}. We {further compare with PPs}, where $v(t)\equiv 0$.
Both active models reduce to PPs under the limits $\drift\to0$ or $\tau\to0$. The limit $\tau\to\infty$ yield{s} a ballistic motion with a constant random velocity set by the initial condition.

{We study the problem of} moving the trap from $\lambda_0=0$ at $t=0$, where the system is in a {nonequilibrium} steady state, to a target position $\lambda(\tf) = \lambdaf$ in a finite time $\tf$.
Due to the frictional resistance of the fluid and changes in the particle's potential energy, such a process generally requires external work input by the controller \cite{jarzynski1997nonequilibrium,crooks1998nonequilibrium},
\begin{equation}
\elabel{def_work1}
  W
    = \int_0^{\tf}\plaind{t}\ \dot{\lambda} \frac{\partial V(x,\lambda)}{\partial \lambda}
    = \int_0^{\tf}\plaind{t}\ \dot{\lambda} k (\lambda - x)\, .
\end{equation} 
We optimize the time-dependent protocol $\lambdaopt$
to minimize the average work input.
We deliberately do not account for the dissipation associated with self-propulsion, as the internal {swimming} mechanism 
is unresolved at RTP/AOUP level, nonuniversal, and hard to measure experimentally 
(to quantify it, one could explicitly model  the {swimming} mechanism \cite{gaspard2017communication,pietzonka2017entropy,speck2018active,speck2022efficiency}).

\textit{Open-loop control.---}First we discuss the open-loop control {problem where no measurements are taken. We minimize the noise-averaged work functional given in \Eref{def_work1} using the Euler-Lagrange formalism, {see \cite{companionpaper} for the explicit derivation. Thereby, the initial condition is drawn from the }nonequilibrium steady state, {where} $\langle x(0)\rangle_{\op} = 0$ and $\langle v(0) \rangle_{\op} = 0$. {We denote averages over trajectories initialized in the steady state by $\ave{\bullet}_{\op}$}. Interestingly, the resulting optimal protocol $\lambda$ turns out to be independent of the self-propulsion, and therefore, \textit{identical} for AOUPs, RTPs and PPs \cite{schmiedl2007optimal}. It} has two symmetric jumps 
at $t=0,\tf$ of length $\lambdaf/(k\tf + 2)$ connected by a linear dragging regime, shown as thin line in Fig.~\ref{fig:work_incs}(b), and its energetic cost is $\Waveopt_{\op} =  \lambdaf^2 /(\tf + 2/k)\geq0$.
Thus, the activity does not decrease the required {average} work.   
These findings are important in experimental settings, where the underlying particle dynamics {and level of activity} may be unknown.

However, the activity influences the \emph{fluctuations} of the work done by the controller.
As we derive in the companion paper \cite{companionpaper}, the variance of the work for PPs is $\Var_{\op} W_{\text{passive}} = 2D  \lambdaf^2 / (\tf + 2/k) $, while for both active models, there
is {an identical} additional contribution 
\begin{align}
& \Var_{\op} W_{\text{active}} - \Var_{\op} W_{\text{passive}} =
\nonumber\\
&\!\!\!= \frac{ 2 \lambdaf^2  \drift^2}{(\frac{\tf}{\tau} + \frac{2}{k\tau})^2} \left[ \frac{\tf}{\tau} + \frac{ \frac{2}{k\tau} + (1 - k\tau) (1 - e^{-\tf/\tau} )}{1 + k\tau}  \right] \!\!
\geq 0\, .\elabel{DeltaVar}
\end{align}
Thus, the variance of the work is always increased by activity, consistent with the picture of an increased effective temperature. The additional variance only vanishes in the trivial (passive) limits:  $\lambdaf\to 0$, $\omega \to 0$, and in the quasistatic limit, $\tf\to \infty$, where $\Var_{\op} W$ vanishes for all models. 

Taken together, our findings suggest that activity {provides no advantage} in the open-loop {problem}: while 
the optimal protocol and minimum mean work remain identical to the passive case, the stochastic work entails larger uncertainty. Furthermore, at the level of first and second moments, the 
(non-)Gaussianity is irrelevant for open-loop control,
though a distinction arises in the higher moments. For example, the positional distribution remains Gaussian for PPs and AOUPs throughout the protocol, while the non-Gaussian distribution of RTPs is generally changed by the dragging \cite{garcia2021run}.

\textit{Closed-loop control (initial measurement).---}We turn to the case where a feedback controller or mesoscopic ``demon'' takes an initial real-time measurement before deciding the protocol. The optimal closed-loop protocol can be obtained by the same procedure as before, now incorporating the (measured) states $x(0)=x_0,v(0)=v_0$ as initial condition. 
We focus on the case where only $v$ is measured, which is obtained by {marginalizing} over $x$. {Notably, a} measurement of $v$ also yields {statistical} information about $x$, {since 
the average of $x(0)$, conditioned on $v(0)=v_0$, is
$\langle x(0) \rangle_{v_0}=  v_0/(k+\tau^{-1}) \not = \langle x(0)\rangle$ \cite{companionpaper,garcia2021run};}
{here we introduced the notation} $\langle \bullet \rangle_{v_0}=\langle \bullet |v(0)\!=\!v_0\rangle$.

As detailed in the companion paper \cite{companionpaper}, the optimal protocol after a $v$-measurement, which again coincides for AOUPs and RTPs, reads
\begin{equation}\elabel{lambdaoptactive}
  \lambdaoptv =
  \left\{
  \begin{array}{l l}
      0, & t=0 \\
      \xaveopt_{v_0} + \frac{d_{v_0}}{k\tf + 2} - \frac{v_0 }{2 k} e^{-t/\tau}\quad & 0 < t < \tf \\
      \lambdaf, & t =\tf
  \end{array}
  \right. \ ,
\end{equation}
with
\begin{align}
  \xaveopt_{v_0} &= \frac{v_0}{k+\tau^{-1}}
  + \frac{d_{v_0}}{\tf + 2/k} t  + \frac{\tau v_0}{2}  (1-e^{-t/\tau})  \ ,\\
d_{v_0} &= \lambdaf - \frac{v_0}{k+\tau^{-1}}
- \frac{\tau v_0}{2}  \left(1-\exp{-\tf/\tau}\right)\, .
\end{align}
The quantity $d_{v_0}$ corresponds to the distance between the target trap position $\lambdaf$ and the expected value of the particle's initial position after {$v_0$} measurement, minus the mean distance covered ``for free'' by the particle's self-propulsion until its orientation decorrelates (last term).

Figures~\ref{fig:work_incs}(a) and (b) show a typical optimal protocol \eref{lambdaoptactive} for some $v_0>0$, $\lambdaf>0$. 
Different from the open-loop case, the jumps at $t=0,\tf$ can be asymmetric and, at intermediate times, the protocol is generally nonlinear. 
Furthermore, the initial jump
\begin{equation}
\elabel{initial_jump}
  \lambda_{v_0}(0^+) - \lambda_{v_0}(0) = \frac{v_0}{k+\tau^{-1}}
  - \frac{{v_0 }}{2 k}+ \frac{d_{v_0}}{k\tf + 2} \,,
\end{equation}
can be in the \textit{opposite} direction to the target position $\lambdaf$.
What seems at first glance even more counterintuitive, these ``reversed'' jumps occur when the initial self-propulsion already pushes the particle towards $\lambdaf$, see Fig.~\ref{fig:work_incs}(b).
The expectation value for the corresponding cumulative work $\int_0^t\plaind{t'}\  \dot{\lambda}(\lambda - \ave{x})$, shown in Fig.~\ref{fig:work_incs}(c), reveals {the reason.} While the initial reversed jump is costly, {it brings the system into a configuration {that enables} subsequent \textit{extraction} of net {work} from the self-propulsion {(negative $\langle \mathrm{d}W\rangle $)}, overcompensating the initial cost.}

The average total work associated with $\lambda_{v_0}$ is
\begin{align}
\elabel{min_work1}
  \Waveopt_{v_0} &=  
    -\frac{k v_0^2}{2(k+\tau^{-1})^2}
    - {\frac{\tau v_0^2}{8}} \left(\!1- \exp{-2\tf/\tau}\! \right) 
    + \frac{d^2_{v_0}}{\tf + 2/k}
    \ ,
\end{align}
which implies that $v$ measurements {enable} work extraction  through two {distinct} physical mechanisms. First, 
measuring $v$ gives average positional information that makes it possible to extract some of the potential energy stored in the initial configuration, {concretely $(k/2)\ave{x(0)}_{v_0}^2$ (first term in \Eref{min_work1})} \footnote{The average potential energy can only be extracted partially, as $(k/2)\ave{x(0)}_{v_0}^2\leq (k/2)\ave{x(0)^2}_{v_0}=\ave{V}_{v_0}$, by the Cauchy-Schwarz inequality}. Second, energy can be extracted  from the self-propulsion: Setting the trap to a position where the particle actively \textit{climbs up} the potential and then moving the trap along the direction of self-prop{elled motion}, one can extract work from the activity until the orientation decorrelates. In contrast to work extracted from potential energy, this operation requires a finite duration, $\tf>0$. Neither mechanism requires an ambient temperature, $D>0$, {unlike} work extraction from potential energy {of PPs} through positional measurements~\cite{abreu2011extracting}. 

\begin{figure}
\centering
\includegraphics[width=\linewidth]{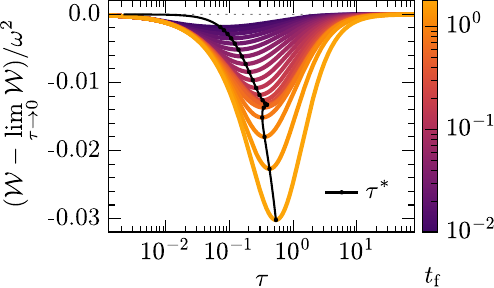}
\caption{The work for closed-loop control averaged over {$v_0$ measurements} exhibits a pronounced minimum for a finite persistence time $\tau$ {(here, $k=1$)}. The $y$-axis shows the reduction in average work compared to the passive limit ($\tau\to0$). {The optimum value $\tau^*$ (black line) is non-monotonic in $\tf$, due to a ``kink" around $\tf=1$, {discussed in more detail in \cite{companionpaper}.}}
}
\label{fig:avg-work}
\end{figure}

Averaging \Eref{min_work1} over the steady state distribution of $v(0)$, {yields the expected mean work}
\begin{align}
\begin{split}
    \mathcal{W} = 
    &- \frac{k \drift^2}{2(k+\tau^{-1})^2} 
    - {\frac{\tau \omega^2}{8}} \left(1- \exp{-2\tf/\tau} \right) 
    \\ 
    &+ \frac{\lambdaf^2 + \left(\frac{2}{k+\tau^{-1}} + \tau (1 - e^{-\tf / \tau})\right)^2\omega^2/4}{\tf + 2/k} \; .
\end{split}
\end{align}
$\mathcal{W}$ has a pronounced minimum at finite $\tau$, Fig.~\ref{fig:avg-work}. 
Thus, the expected cost to translate an active particle is lower than for a {PP} ($\tau \to 0$), and, remarkably, also a ballistic particle ($\tau \to \infty$). In both these limits the work $\mathcal{W}$ reduces to $\frac{\lambdaf^2}{\tf + 2/k}$. Therefore, a finite persistence time, i.e., the presence of nonequilibrium fluctuations, poses an advantage for closed-loop control.

\begin{figure*}
\includegraphics[width=\textwidth]{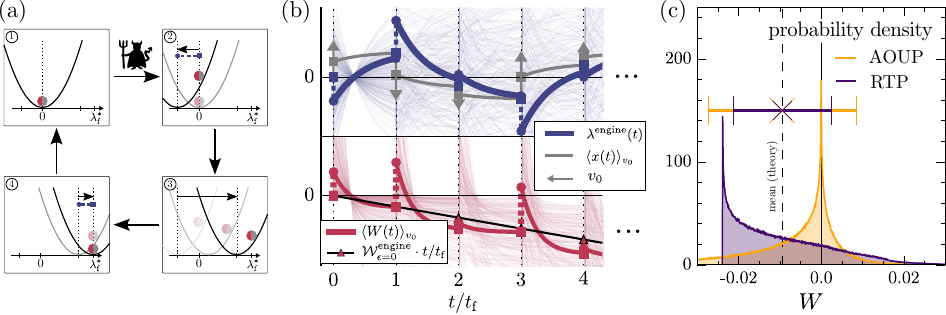}
\caption{Active information engine: (a) iterated protocol $\lambdaengine$ determined upon an initial self-propulsion measurement $v_0$ in each iteration, 
(b) single realization of engine protocol (thick blue), corresponding noise-averaged particle position (grey), and cumulative work (thick red). The arrows indicate the measured $v_0$ values. Thin lines show trajectories for other {stochastic realizations} ($\omega = 1.0, \tau = 0.5, \tf = 10, k = 0.5$).
(c) Work distribution per iteration for AOUP and RTP from simulations ($\omega = 0.2, \tau = 1, \tf = 10, k = 2$). While the mean ($\times$) is identical for AOUP and RTP, matching the theory (dashed line), AOUPs have a larger standard deviation (vertical bars). For RTPs, the unique mode of the distribution is below the mean, while it is close to zero for AOUPs.
}
\label{fig:engine}
\end{figure*}

\textit{Active information engine.---}A notable potential application of optimal protocols are optimized periodic machines. Based on our results so far, one can readily construct a minimal, optimal information engine that harvests energy from activity using self-propulsion measurements, as follows. From Eqs.~\eref{lambdaoptactive}-\eref{min_work1}, we find that the work $\Waveopt_{v_0}$ is minimized, i.e., work extraction is maximized, if the trap is set to {$\lambdaf^* = \frac{v_0}{k+\tau^{-1}} + \frac12\tau v_0(1-e^{-\tf/\tau}) $}. 
For this optimal target displacement $\lambdaf^*$, the final trap position coincides with the average particle position at the end of the protocol,
so that the protocol extracts any average potential energy {available at $t=\tf$}.
{Setting $\lambdaf=\lambdaf^*$} further yields $\ave{W}_{v_0} \leq 0$ for all $v_0$, and $d_{v_0}=0$. The corresponding optimal protocol reads
\begin{equation}\elabel{lambdaengine}
  \lambdaengine =
  \left\{
 \begin{array}{l l}
     0, & t=0 \\
    \frac{\tau v_0}{2}\left(\frac{3 + k\tau}{1+k\tau} - \frac{1 + k\tau}{k\tau} e^{-t/\tau} \right) \quad & 0 < t < \tf
    \\
    \frac{\tau v_0}{2}\left(\frac{3 + k\tau}{1+k\tau} - e^{-\tf/\tau} \right) & t =\tf
  \end{array}
  \right. \ .
\end{equation} 
Iterating $\lambdaengine$ yields an {information} engine with period $\tf$, visualized in Fig.~\ref{fig:engine}. {Over long timescales, the trap center undergoes diffusive motion. This can however be prevented by reset events that do not asymptotically affect efficiency} \footnote{{The  diffusive delocalization of the trap center can be prevented by incorporating a recovery step of duration $t_\mathrm{rec}$ that resets the system to the origin after any $N$ cycles, yielding a process that is cyclic on the level of average quantities with 
period length $N  t_\mathrm{f}+t_\mathrm{rec}$. If $t_\mathrm{rec}$ is chosen to scale sublinearly with $N$, the corrections to both efficiency and power also scale sublinearly. This implies that all our results derived for the operation without recovery steps remain asymptotically valid for large $N$. For further details see \cite{companionpaper}.}}.

The motor builds up nontrivial correlations between iterations, so the system does not reach a steady state. 
However, {the orientational dynamics is not affected by the trap, and
the trap position automatically coincides with the average particle position at the end of the protocol. Therefore, assuming that $\tf$ is large compared to typical intrinsic relaxation times}~\footnote{{We can assume that the particle has {effectively reached a steady state} by the end of the iteration if its duration $\tf$ is greater than the {intrinsic relaxation} timescales {of $v$ and $x$,} $\tau$ and $k^{-1}$, {so that we can apply} \Eref{min_work1}.}
}, 
{we can still calculate the average work extraction per iteration from \Eref{min_work1}, which yields
$ {\frac{k }{2}\omega^2 v_0 (k+\tau^{-1})^{-2}}
- \frac{1}{8} \omega^2 \tau \left(1- e^{-2\tf/\tau} \right) 
$
for AOUPs and RTPs.
This has no optimum $\tau$, and reaches its maximum in the quasistatic regime. 
Simulating this engine,} we find that, while the average work is identical for AOUPs and RTPs, the work distributions differ significantly, {and most notably are always broader for AOUPs, }Fig.~\ref{fig:engine}(c).

Finally, we address the effect of measurement uncertainty and 
the thermodynamic cost of information  acquisition~\cite{sagawa2010generalized,kullback2013topics,cao2009thermodynamics,parrondo2015thermodynamics}, which has not been included in previous studies of active information engines \cite{cocconi2023optimal}. 
{The uncertainty of the $v$ measurement lowers the extractable work.
Assuming a Gaussian-distributed {measurement} error $\epsilon$,} we show in \cite{companionpaper} that the net extracted work has an additional cost term $\propto \epsilon^2$, leading to a total average work per iteration of 
\begin{equation}\elabel{Wengine}
\Weng = 
(\epsilon^2-\omega^2) \left( \frac{k/2}{(k+\tau^{-1})^2}
+ \frac{\tau}{8 } \left(1- e^{-2\tf/\tau} \right) \right) \,.
\end{equation}
{Thus, }work extraction is {still} possible for all finite $k, \tau, \tf$, provided that $\epsilon < |\omega|$.

The information efficiency $\eta$ of the engine is obtained by dividing the mean extracted work per iteration by the information acquisition cost. {Henceforth}, we assume an isothermal process, where the demon operates at temperature ${D}=D_v$. Under this assumption, the fundamental limit to the information acquisition cost for an individual measurement is given by the relative entropy and, when averaged over multiple iterations, by 
the mutual information $\mathcal{I}$ between the true value $\vgt$ and the measurement outcome $\vm$, scaled by the temperature~\cite{sagawa2010generalized,sagawa2008second,cao2009thermodynamics,abreu2011extracting,horowitz2010nonequilibrium}. {Thus,}
\begin{align}
\elabel{def_efficiency}
    \eta = \frac{-\Weng}{D_v\,\mathcal{I}} \,.
\end{align}

{For AOUPs, the mutual information is }\cite{companionpaper}
\begin{align}\label{I:AOUP}
\mathcal{I}_\mathrm{AOUP }[\vm;\vgt]=
\frac{1}{2} \ln \left( 1 +  \frac{{\drift}^2}{\epsilon^2}\right)\, ,
\end{align}
{which} vanishes for $\epsilon \to \infty$, where measurements become useless, and diverges for $\epsilon \to 0$. This divergence is not surprising, in view of the infinite information content in a continuous variable, $v_0 \in \mathbb{R}$ \cite{PolyanskiyWu2014,Gray2011}. For RTPs, 
we find in the regime of small errors $\epsilon\ll\drift$, 
\begin{align} \label{I:RTP}
\mathcal{I}_\mathrm{RTP}[\vm;\vgt]= \ln (2)\, ,
\end{align}
{i.e.,} the measurement yields exactly \textit{one bit} of information. Comparing \eqref{I:AOUP} and \eqref{I:RTP}  indicates that, at least 
in the regime of interest of modest {$\epsilon$}, the cost of information acquisition is generally {lower} for RTPs than AOUPs. 
In fact, RTPs have the smallest information cost (one bit) of \textit{any} possible active particle model.
Recalling that the extractable work is independent of the model of self-propulsion, we conclude that the information efficiency $\eta$ is highest for RTPs.

Together with \Erefs{Wengine} and \eqref{I:RTP}, this gives a universal upper bound to the information efficiency,
\begin{align}
    \eta \leq \eta_\mathrm{max} =  {1}/{[4\ln(2)]}\, .
\end{align}
The engine is most efficient when run with an RTP at low frequency $1/\tf$, with low error $\epsilon$, and $\tau=k^{-1}$
\footnote{Taking into account the underlying thermodynamic cost of the self-propulsion as a background dissipation term 
results in a maximum efficiency at some finite frequency $1/\tf$. {For example, assuming a constant rate $\mathcal{A}$ of ATP depletion of the self-propulsion mechanism, the efficiency {scales like} $\eta \propto \frac{\Weng}{D_v\,\mathcal{I} +\mathcal{A}\, \tf}$.}}. Even for finite $\tf$, the efficiency $\eta$ admits a maximum at a finite persistence time $\tau$. 

\textit{Conclusions.---}We have generalized a canonical optimization problem to the case of active particles.
The open-loop protocol is not affected by activity, but fluctuations of the work are generally increased. In contrast, closed-loop protocols can extract energy from self-propulsion measurements by jumps away from the target position. Surprisingly, a finite persistence time is beneficial compared to ``infinitely persistent'' (ballistic) particles. 
We have further derived an optimal engine to harvest work from the self-propulsion measurements by simple shifts of the trap. Such an engine is most efficient and most precise for RTPs{ benefiting} from non-Gaussianity. We found a universal upper {information} efficiency bound for this engine design of $1/[4\ln(2)]\approx 0.36$. 
Higher efficiencies are presumably achievable with more {complex} machines, e.g., by dynamic adjustment of trap stiffness.

{Our results for AOUPs readily generalize to higher dimensions, and it would be compelling to explore other active models in higher dimensions too.}
To {address} more complicated processes, recent machine learning algorithms offer a promising approach \cite{whitelam2023demon,engel2023optimal,loos2023universal}, potentially enabling the study of collective active systems.
We hope that our work will stimulate experimental investigations into optimal control of optically trapped bacteria \cite{ashkin1987optical} or synthetic active particles \cite{buttinoni2022active,goerlich2022harvesting}.

\begin{acknowledgments}
We thank Robert Jack and \'Edgar Rold\'an for insightful discussions.
R.G.-M. acknowledges support from a St John's College Research Fellowship, Cambridge.
J.S. acknowledges funding through the UK Engineering and Physical Sciences Research Council (Grant number 2602536). 
S.L. acknowledges funding by UK Research and Innovation (UKRI) under the UK government’s Horizon Europe funding Guarantee (grant number EP/X031926/1), through the Walter Benjamin Stipendium (Project No. 498288081) from the 
Deutsche Forschungsgemeinschaft (DFG), and from Corpus Christi College Cambridge. 
\end{acknowledgments}

\bibliography{bib.bib}

\end{document}